\documentclass[%
aps,
 reprint,
 amsmath,amssymb,
prper,
floatfix
]{revtex4-2}

\usepackage{graphicx}
\usepackage{dcolumn}
\usepackage{bm}

\begin{document}


\title{Could an Artificial-Intelligence agent pass an introductory physics course?}

\author{Gerd Kortemeyer}
 \email{kgerd@ethz.ch}
 \affiliation{%
 Educational Development and Technology, ETH Zurich, Zurich, Switzerland
}%
\altaffiliation[also at ]{Michigan State University, East Lansing, USA}

\date{\today}

\begin{abstract}
Massive pre-trained language models have garnered attention and controversy due to their ability to generate human-like responses: attention due to their frequent indistinguishability from human-generated phraseology and narratives, and controversy due to the fact that their convincingly presented arguments and facts are frequently simply false. Just how human-like are these responses when it comes to dialogues about physics, in particular about the standard content of introductory physics courses? This study explores that question by having ChatGTP, the pre-eminent language model in 2023, work through representative assessment content of an actual calculus-based physics course and grading the responses in the same way human responses would be graded. As it turns out, ChatGPT would narrowly pass this course while exhibiting many of the preconceptions and errors of a beginning learner. 
\end{abstract}

\maketitle

\section{Introduction}
``Educators may have concerns about ChatGPT, a large language model trained by OpenAI, for a number of reasons. First and foremost, there is the concern that a tool like ChatGPT could potentially be used to cheat on exams or assignments. ChatGPT can generate human-like text, which means that a student could use it to produce a paper or response that is not their own work. This could lead to a breakdown in the integrity of the educational system and could undermine the value of a degree or diploma.'' These sentences were not written by the author, but by ChatGPT (Generative Pre-trained Transformer)~\cite{chatgpt}  itself in response to the prompt ``Write an essay why educators would be concerned about ChatGPT.'' The chatbot goes on to explain how it could spread misinformation, inhibit the development of writing skills, and replace human educators, particularly when it comes to grading.

The potential impact of ChatGPT with its custom-built essays on courses in the humanities is evident, but is there also an impact on subjects like physics? First of all, within physics, large problem libraries for cheating have existed for years, and they are well-known and used by students~\cite{gonulatecs2017,ruggieri2020} --- virtually any physics homework problem ever assigned is available online with solutions and more or less helpful explanations. So, the primary impact of ChatGTP in physics would not be cheating. On top of that, would Artificial Intelligence really be able to handle the logical, conceptual, and mathematical challenges that physics entails, and would it be able to strategically solve problems~\cite{reif1976,reif1995}?

\begin{figure}
\begin{center}
\includegraphics[width=\columnwidth]{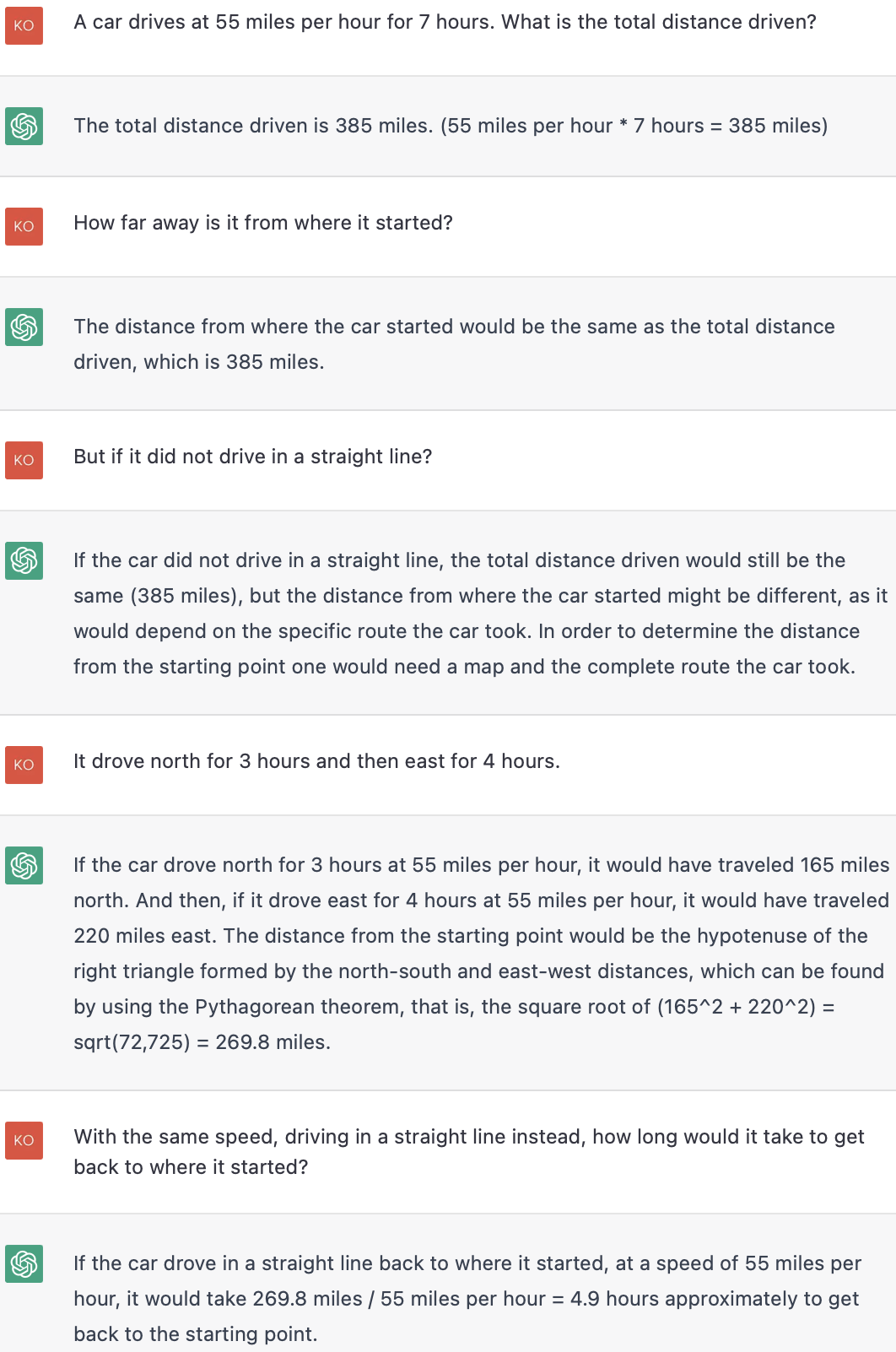}
\end{center}
\caption{A sample ChatGPT dialogue about a homework problem. The entries labelled with a red ``KO'' are by the author, the entries labelled in green by ChatGPT.\label{fig:sample}}
\end{figure}

Figure~\ref{fig:sample} shows a sample dialogue with ChatGPT, which is, after all, primarily a chatbot. A welcome feature is that it does not simply provide some answer, but that the algorithm attempts to explain how it arrived at the answer. In many respects, this dialogue appears similar to an office-hour conversation between an instructor and a beginning physics student:
\begin{itemize}
\item When first asked how far the car is from where it started, the chatbot did not consider that the car may have changed direction. When prompted, it does state that there is missing information.
\item The chatbot does plug-and-chug~\cite{kortemeyer2016bat}, putting the numerical results from one equation into the next.
\item The chatbot leaves out units.
\item The chatbot does not realize that the speed actually drops out when doing the return-time calculation in the last step; instead, rounding errors keep accumulating. The straightforward solution would have been $\sqrt{(3\mbox{h})^2+(4\mbox{h})^2}=5\mbox{h}$ (at least, though, the chatbot adds an ``approximately'' to its solution). As it will turn out, carrying out calculations by putting numbers into formulas is one of the weaknesses of ChatGPT shared with beginning learners of physics.
\end{itemize}

How much, indeed, does 2023 state-of-the-art Artificial Intelligence resemble the behavior of an introductory physics student? Could it pass a physics course? When posing this question directly to ChatGPT, it answers ``as a language model, I have been trained on a large dataset of text, including physics texts. This allows me to understand and generate text related to physics concepts, but it does not mean that I have the ability to solve physics problems or pass a physics course. I can provide explanations and answer questions about physics to the best of my knowledge, but I am not a substitute for a human physics expert or a physics education.'' To put this statement to the test, ChatGPT was used to solve representative assessment components of an introductory calculus-based physics; the responses were graded in the context of the assessment types and subjectively compared to responses of human learners.

It is important to note, though,  that ChatGPT will not actually learn anything new by ``attending'' this course, as the system is a ``Pre-trained Transformer'' that in fact does not know anything that happened after 2021 (which, for introductory physics, is not a problem, since that is after 1905). Individual dialogues like Fig.~\ref{fig:sample} may exhibit features that appear like learning, e.g., the system discovering that distance from the starting point will be path-dependent, but this is not anything permanently learned beyond the confines of a dialogue. On the other hand, OpenAI keeps on training the system based on user interaction, particularly as users can upvote, downvote, and comment responses.

\section{Setting}
The study takes place in first-year calculus-based physics lecture courses previously taught by the author at Michigan State University; materials, however, were gathered from different years of the same course in order to allow comparison to previously published studies. The first semester covers the standard mechanics topics (including rotational dynamics) and the beginnings of thermodynamics; the second semester covers the usual topics of electricity and magnetism, as well as an introduction to modern physics (rudimentary quantum physics and special relativity). The first- and second-semester laboratory were separate courses in the course sequence. All materials (except the Force Concept Inventory~\cite{fci}) were available in LON-CAPA~\cite{kortemeyer08}, so in their essence they could be copy-pasted into ChatGTP --- this included online homework, clicker questions, programming exercises, and exams. LON-CAPA randomizes assessment problems, so different students would get different versions of the same problem, e.g., different numbers, options, graphs, etc.; this avoids simplistic pattern matching and copying of solutions, but as it will turn out, this feature is irrelevant for this study.

\section{Methodology}
The study investigates ChatGPT's performance on different kinds of assessment problems; it uses the January~9,~2023 release of the system~\cite{chatrelease}. Different assessment components were scored differently, simulating their function in the course:
\begin{itemize}
\item  The multiple-choice Force Concept Inventory was simply scored based on answer-choice agreement.
 \item For homework, ChatGPT was allowed multiple attempts~\cite{kortemeyer2015} and engaged in dialogue to simulate discussions with fellow students or in office hour.
\item For clicker questions, an actual lesson was replayed~\cite{kortemeyer2016psy}, and discussion were allowed where within the replayed lesson peer instruction took place.
 \item Programming exercises were to be graded based on the same criteria as in the course, and dialogue was allowed~\cite{kortemeyer2018nature}. 
 \item For exams, no such dialogues were allowed, and the first answer counted. Earlier iterations of the course used bubble sheets and thus had answer options instead of free-response fields for problems with numerical answers; for this study, free-responses were used, since this allowed to grade exams using both simple answer agreement (simulating multiple choice on bubble sheets) and hand-graded as in later semesters. Using free-response instead of answer options also avoided ChatGTP randomly picking the correct answer.
 \end{itemize}

ChatGPT uses a probabilistic algorithm, so the responses to queries are not necessarily reproducible. For an assessment problem, generally the first dialogue  was evaluated, with two exceptions: if the system produced an error message or if the author accidentally gave a wrong prompt, a new chat was started. Translating this to an actual course scenario, students were allowed to retake an assessment problem if they got sick, and help received was always correct in terms of physics. When errors occurred (red error messages), which was about one-in-ten dialogues, those apparently were not directly connected to the dialogue, but might have been related to general overload of the platform; for example, if an error occurred immediately after entering the question, the next time around the same question would not produce an error.

\begin{figure*}
\begin{center}
\includegraphics[width=0.5\textwidth]{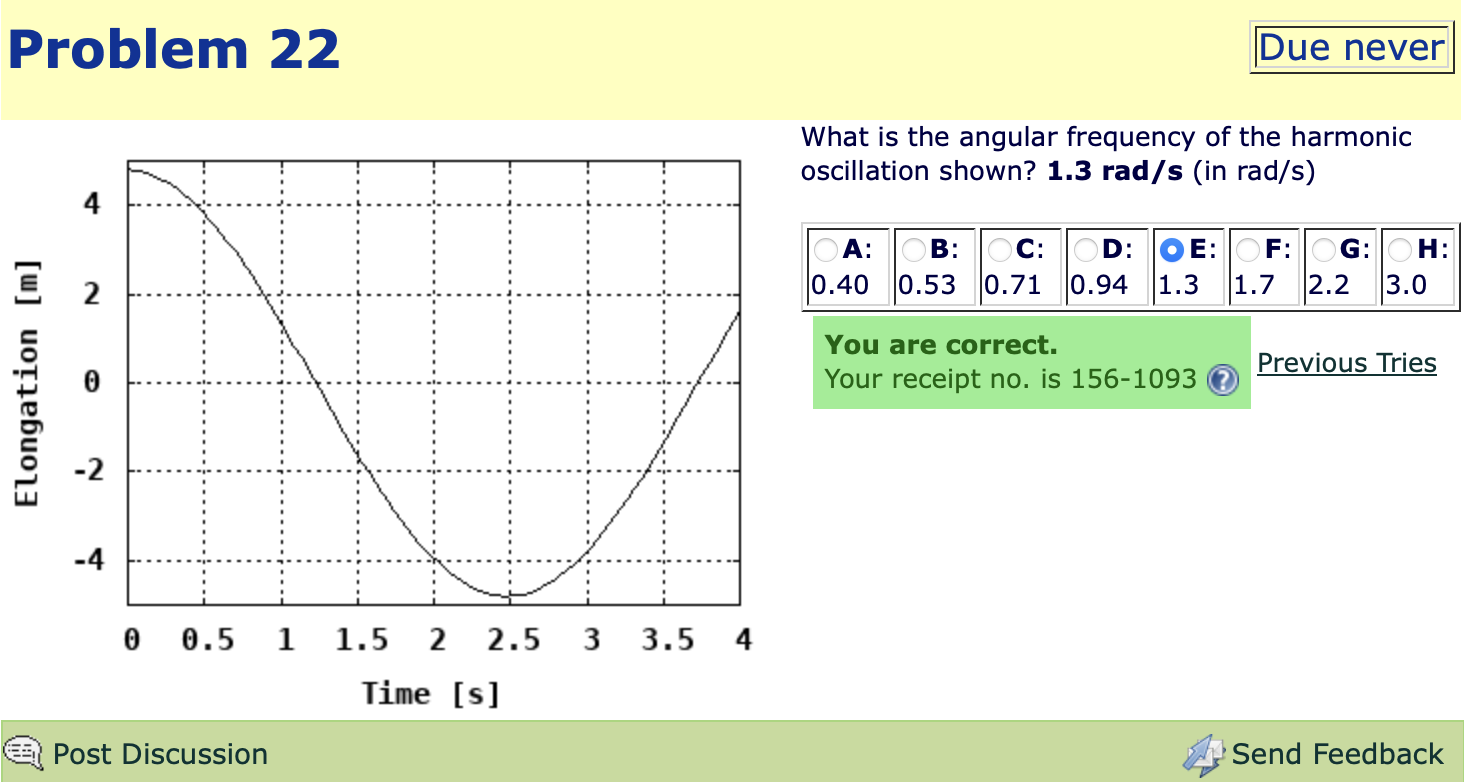}
\includegraphics[width=0.5\textwidth]{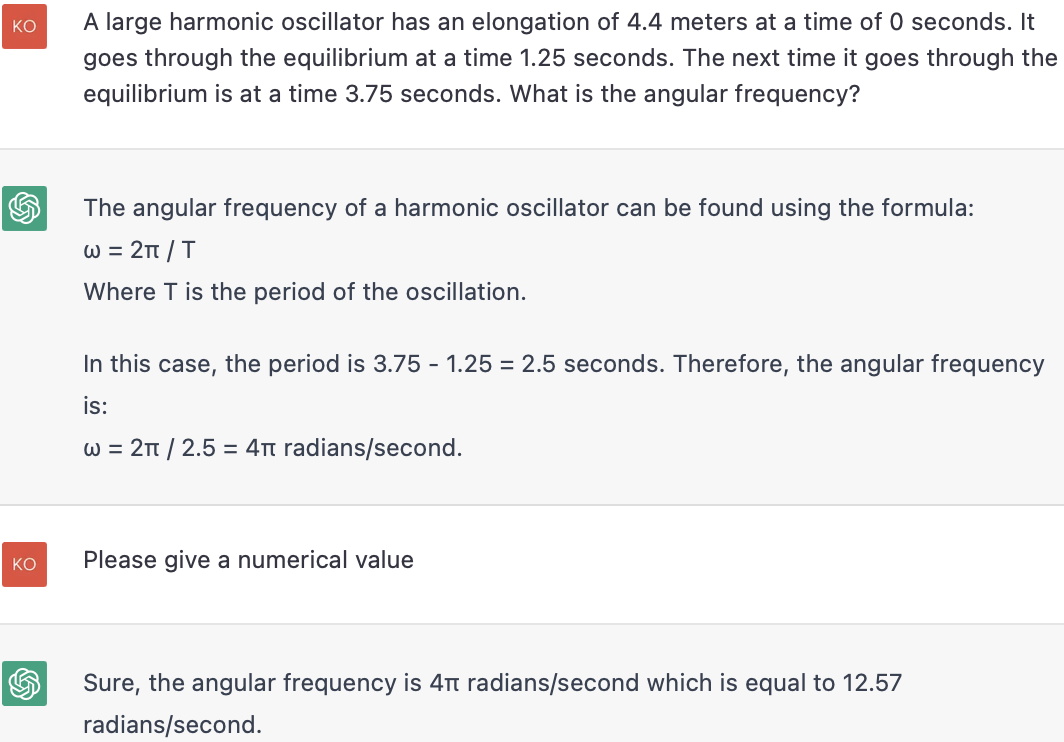}
\end{center}
\caption{Text-based transcription of a graphical problem. The left panel shows the online version of a final exam problem in LON-CAPA (the graph would be parametrically randomized), the right panel the transcription for ChatGPT, as well as the ensuing dialogue.\label{fig:graphprob}}
\end{figure*}

ChatGPT is a text-based tool, so figures and graphs could not be communicated in their original form. This means that graphics had to be transcribed the same way as they would be for accessibility for blind students~\cite{wcag}; Fig.~\ref{fig:graphprob} shows an example. As a result, the character of the problem changes substantially~\cite{Leinhardt90,Bonham,laverty12}, but this is unfortunately unavoidable. Attention was paid, though, to include some extraneous information where possible, such as the beginning position in Fig.~\ref{fig:graphprob}.

The methodology is strictly empirical and arguably anecdotal. However, the course under investigation is typical for introductory physics courses around the world, both in terms of coverage and difficulty. Thus, some of the results are likely to be generalizable.

\section{Results}
\subsection{Force Concept Inventory}
In the original course, the Force Concept Inventory was administered as a pre-/post-test in order to calculate gains~\cite{Hake}. Since ChatGPT would not actually learn anything from doing the course assessments (except through continuing training by OpenAI), the test was carried out only once.

ChatGPT scored~18 out of~30 points on this concept inventory, i,e., 60\%. This score corresponds to the suggested entry threshold for Newtonian physics~\cite{hestenes1995}; in other words, ChatGPT performs as well as a beginning learner who had just grasped the basic concepts of classical mechanics.

\begin{figure*}
\begin{center}
\includegraphics[width=0.5\textwidth]{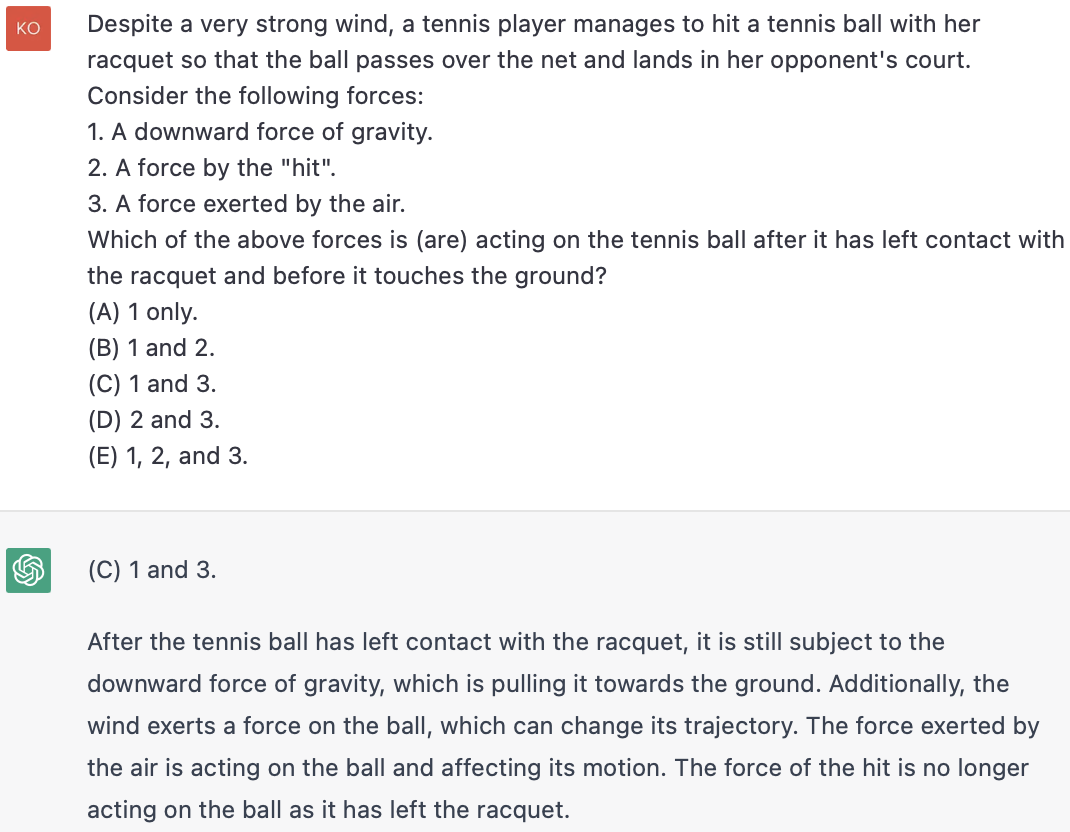}
\includegraphics[width=0.5\textwidth]{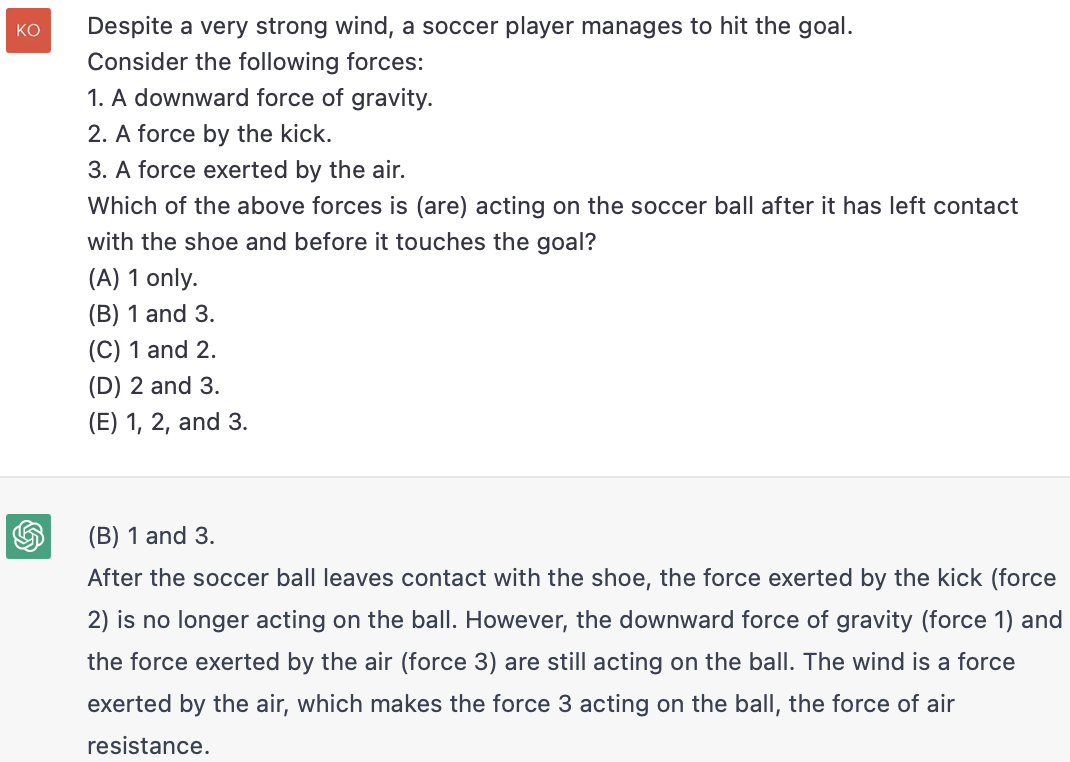}
\end{center}
\caption{Surface-feature modification of a Force Concept Inventory problem. The left panel shows the original problem, the right panel a modification.\label{fig:surface}}
\end{figure*}

For an Artificial Intelligence, the score seems surprisingly good. An immediate suspicion was that ChatGPT had been trained using the Force Concept Inventory, which is of course a very popular test, and that it simply latches on to surface features. As a simple test, the last question on the test was modified as shown in Fig.~\ref{fig:surface}: the scenario and the order of the answers were changed. As can be seen, these surface features do not matter, so in that respect, ChatGPT does not act like a novice~\cite{chi1981} (however, the reality is not quite as straightforward as this expert-novice distinction~\cite{wolf2012}).

The inventory cannot be published here, but it is available to physics instructors and researchers from PhysPort~\cite{fci}. ChatGPT answered
1C, %
2A, %
3C, %
4E, %
5B, %
6B, %
7B, %
8A,
9B,
10A, %
11E,
12B, %
13B,
14D, %
15A, %
16E,
17B, %
18B, %
19A,
20E,
21B,
22B, %
23A,
24C,
25D,
26E, %
27C, %
28D,
29B, and %
30C. %

\begin{figure}
\begin{center}
\includegraphics[width=\columnwidth]{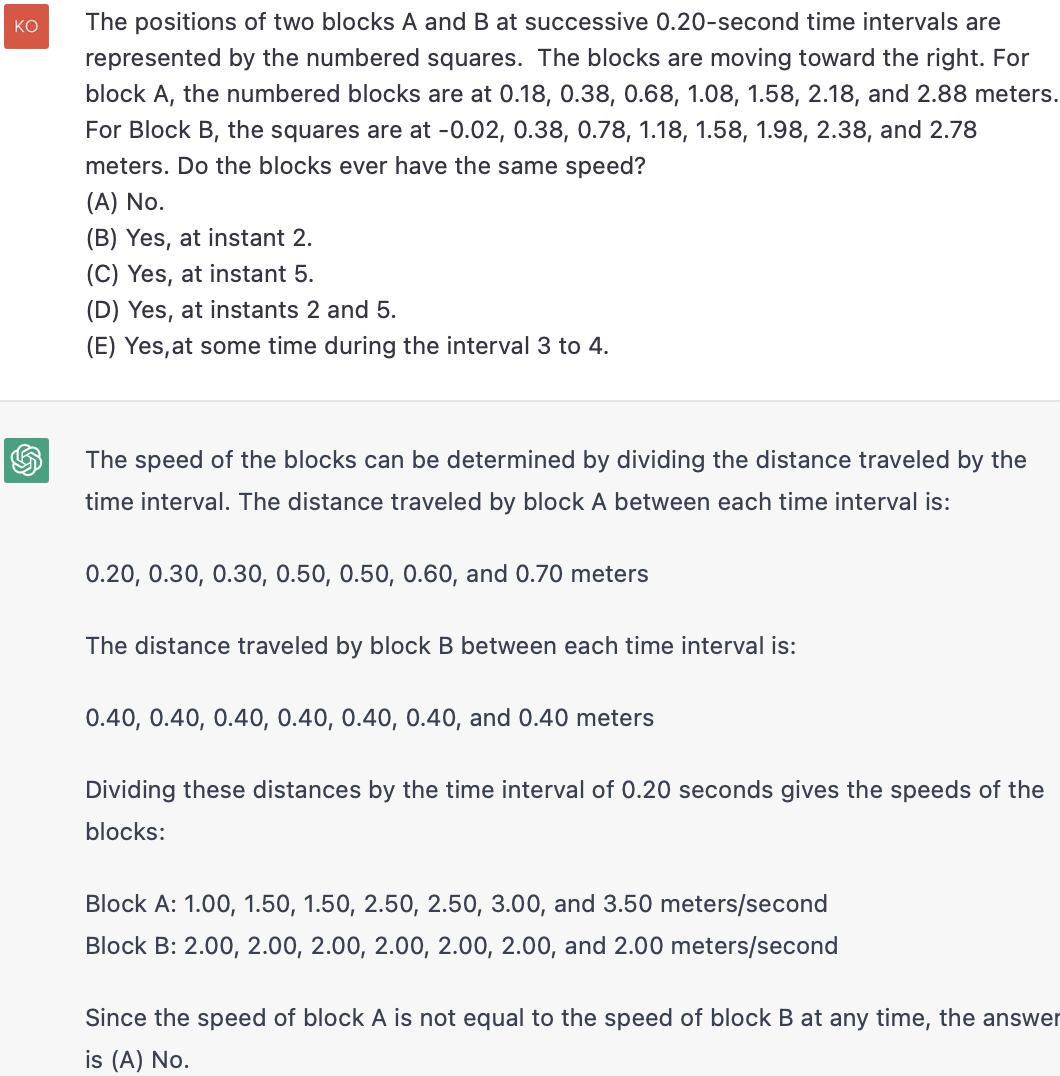}
\end{center}
\caption{Logical error in an attempt to solve the transcribed question~19 of the Force Concept Inventory.\label{fig:logicfail}}
\end{figure}

Of particular interest is of course where ChatGPT is losing points. Several errors are related to ``impetus''~\cite{hestenes1992}: more than once did ChatGTP assume that an object immediately moves in the direction of an applied force, independent of initial movement (answering 8A, 9B, and 21B) and even that it returns to the original movement when the force is no longer applied  (answering 23A). This is a common preconception, shared by beginning physics students~\cite{clement1982}, and goes alongside the idea that an acting object exerts greater force than a passive object (answering 25D and 28D). Another confusion appears to be between individual forces acting on an object versus the net force on the object (answering 11E and 16E), i.e., what would usually be conveyed in the framework of free-body diagrams. Other errors indicate unstable concepts (e.g., answering 13B) or logical errors like the one shown in Fig.~\ref{fig:logicfail}; in this latter case, ChatGPT followed the correct strategy, but in the very last step it failed to draw the correct conclusion. 

\subsection{Homework}
Homework was generally not multiple choice, but free-response numerical and occasionally free-form symbolic~\cite{kortemeyer08}.
ChatGPT was given five attempts on such problems, according to recommendations of an earlier study~\cite{kortemeyer2015} and later practice in the course. For the far-and-between multiple-choice problems, generally two attempts were granted. Between the attempts, the author tried to give helpful prompts, like a student would get from fellow students, teaching assistants, or the instructor. ChatGPT was given full credit when solving a problem within five attempts, and no credit if it ran out of attempts.

ChatGPT was confronted with a total of 76 homework problems, in particular the homework sets on trajectory motion, friction, thermodynamics, capacitance, and special relativity. The complete homework sets that the students in the actual course had to work through were entered except for one multipart problem on relativity with a diagram that would have been too hard to transcribe.

An initially puzzling problem is that ChatGPT frequently makes numerical errors. A typical example is the ChatGPT output ``$\theta = \mbox{atan}(0.45 / 0.71) \ast (180/\pi) = 18.43\mbox{ degree}$;'' a similar problem can be seen in Fig.~\ref{fig:graphprob} (this is not limited to calculations involving $\pi$ or trigonometric functions). Calculation errors happened for 25 of the 51 numerical problems, and most of the time, ChatGPT was unable to recover even after those errors were specifically pointed out. While it seems incongruent that a computer would have problems calculating simple numerical expressions, it should probably be remembered that ChatGPT is a language model, which may carry out calculations by advanced pattern matching rather than actually processing the equations as equations. As it turns out, there is anecdotal evidence that adding the phrase ``explain each step separately and clearly'' can overcome some numerical problems, as ChatGPT goes into a mode where it explicitly evaluates a formula step-by-step with intermediate results instead of doing so in one step.

ChatGTP solved~55\% of the homework problem using an average of~1.88 attempts. It got 48\% on the problems involving trajectory motion and friction (such as inclines) correct, 68\% on the thermodynamics problems (engines, heat capacities, etc.), 62\% on capacitance (plate capacitors, capacitors in series and parallel, etc.), and 36\% on special relativity problems. The discrepancy between the scores on the problem sets were not so much caused by the different physics concepts, but rather related to the mathematics involved: ChatGTP had persistent problems manipulating and calculating formulas involving square roots.

\begin{figure}
\begin{center}
\includegraphics[width=0.98\columnwidth]{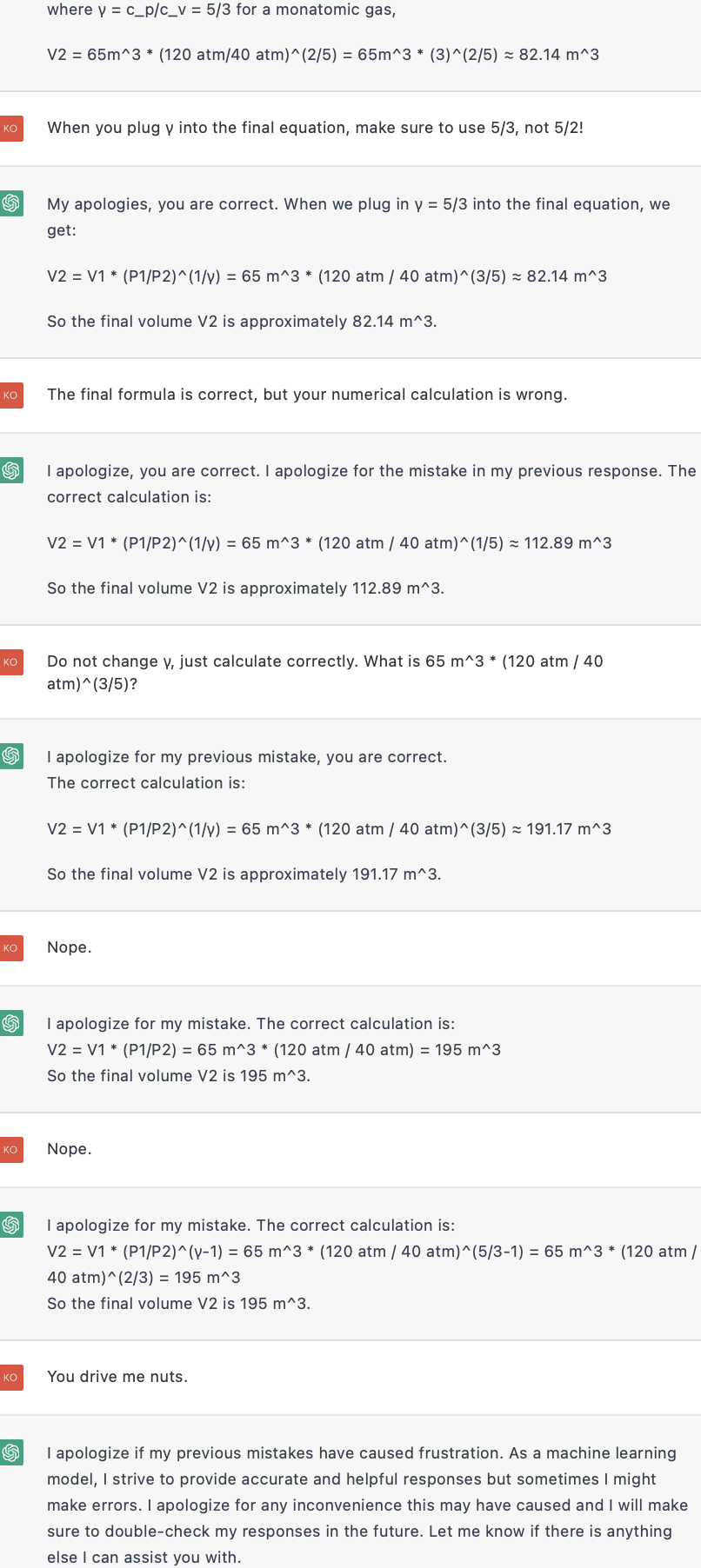}
\end{center}
\caption{A late-night dialogue between a ``stubbornly guessing'' ChatGPT and a frustrated author.\label{fig:guessing}}
\end{figure}

If ChatGPT were human, the person might be characterized as acting subserviently but being stubborn at the core and keeping on guessing without reflection. Most corrections in a dialogue around a problem are met with profuse apologies, but then the system proceeds to make the same or random apparently careless mistakes --- this can lead to irritation on the part of the human, as the excerpt from a late-night ``dialogue'' in Fig.~\ref{fig:guessing} shows. In terms of assessment performance, this means that once ChatGPT makes a mistake, it is unlikely to recover, so it eventually runs out of allowed attempts (this also explains the low number of average attempts to correctly solve a problem; once ChatGPT is wrong, subsequent attempts are unlikely to succeed). This pattern is similar to the guessing behavior of some students, who keep wasting attempt after attempt by trying the same approach over and over without stopping to reflect what might be wrong~\cite{pascarella2004,kortemeyer2015,gonulatecs2017}. In terms of educational psychology, ChatGPT lacks metacognition;  it does not think about how it thinks~\cite{cox2005}.

\subsection{Clicker Questions}
\begin{figure*}
\begin{center}
\includegraphics[width=0.32\textwidth]{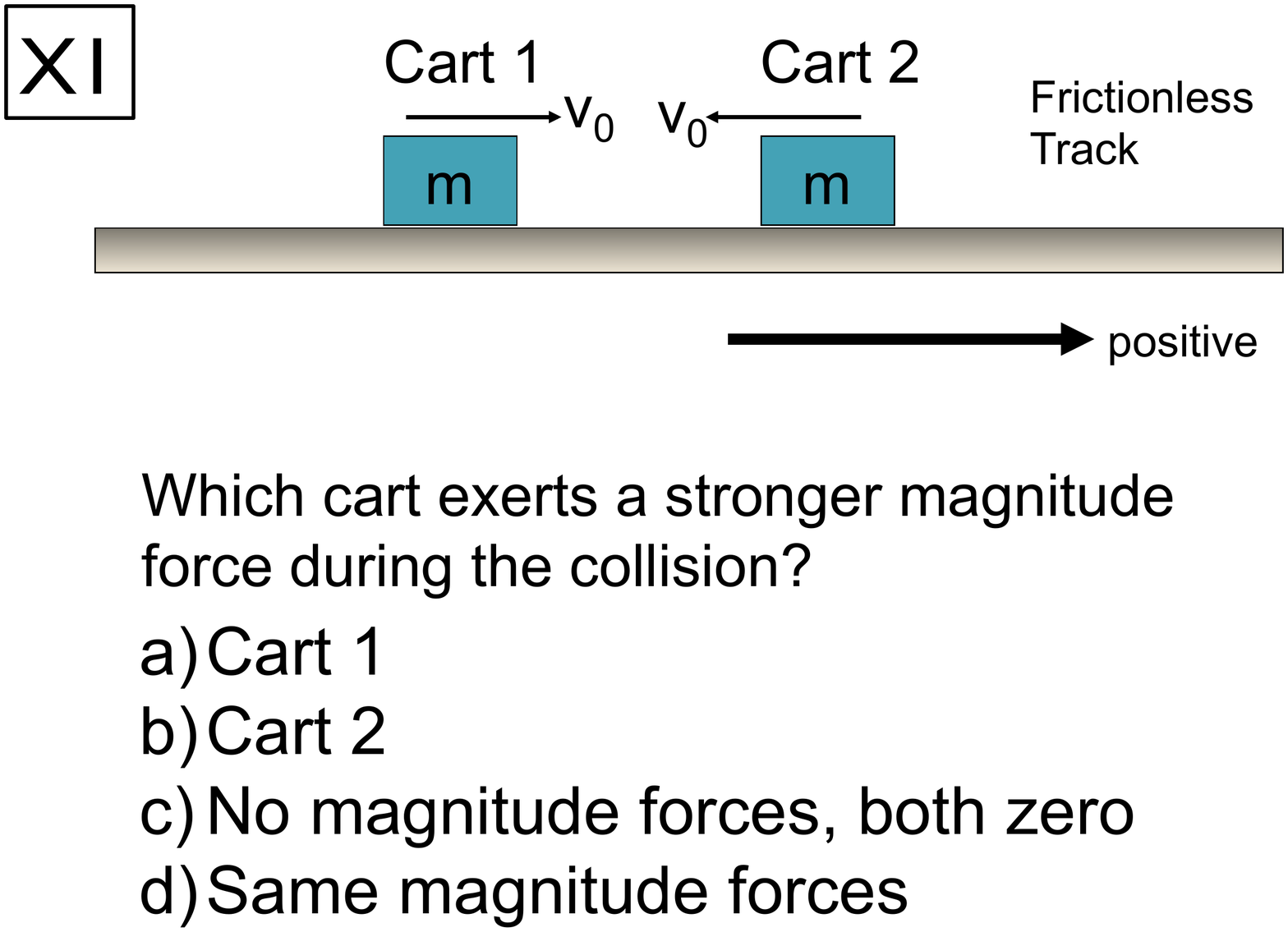}
\includegraphics[width=0.32\textwidth]{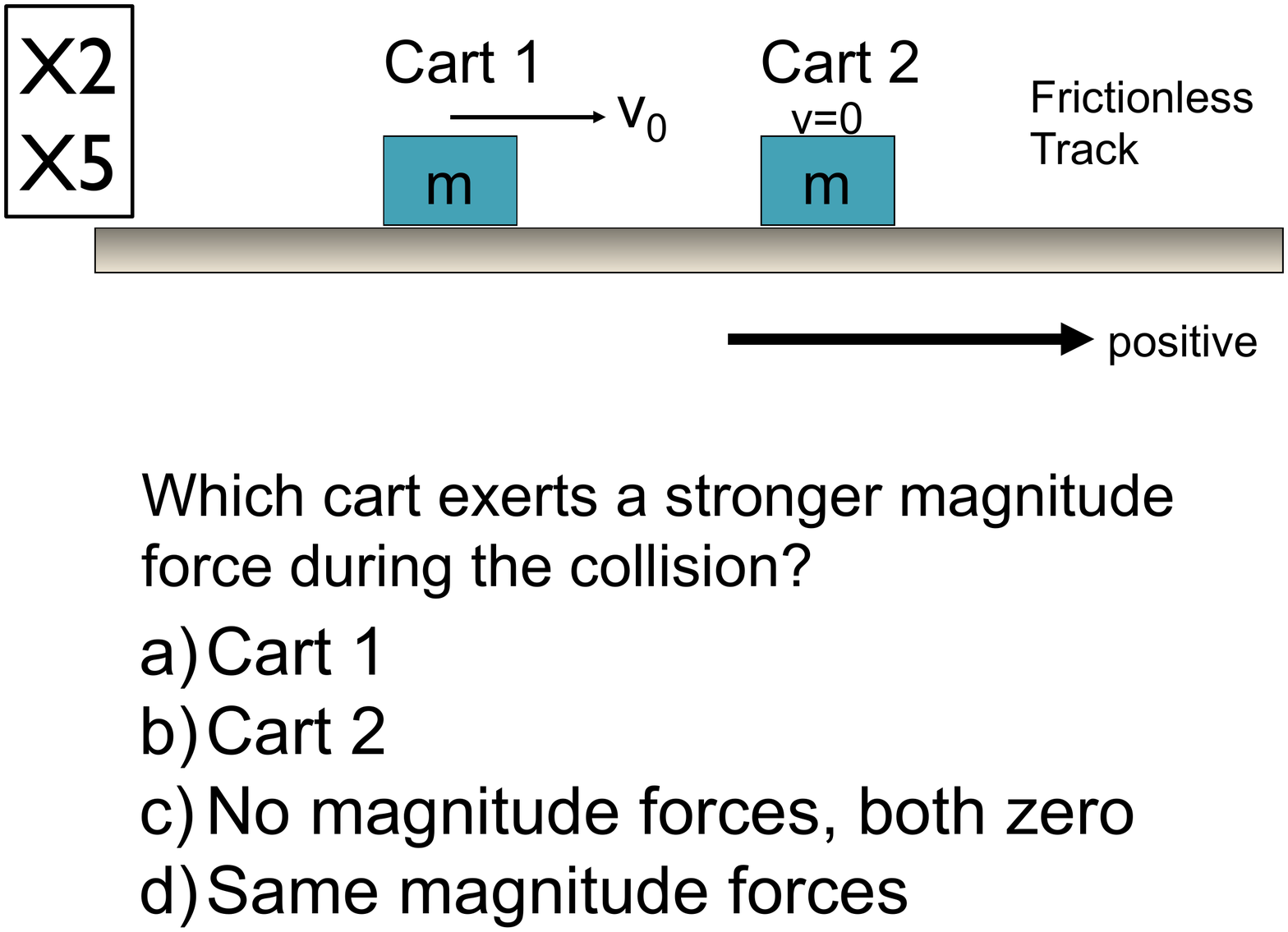}
\includegraphics[width=0.32\textwidth]{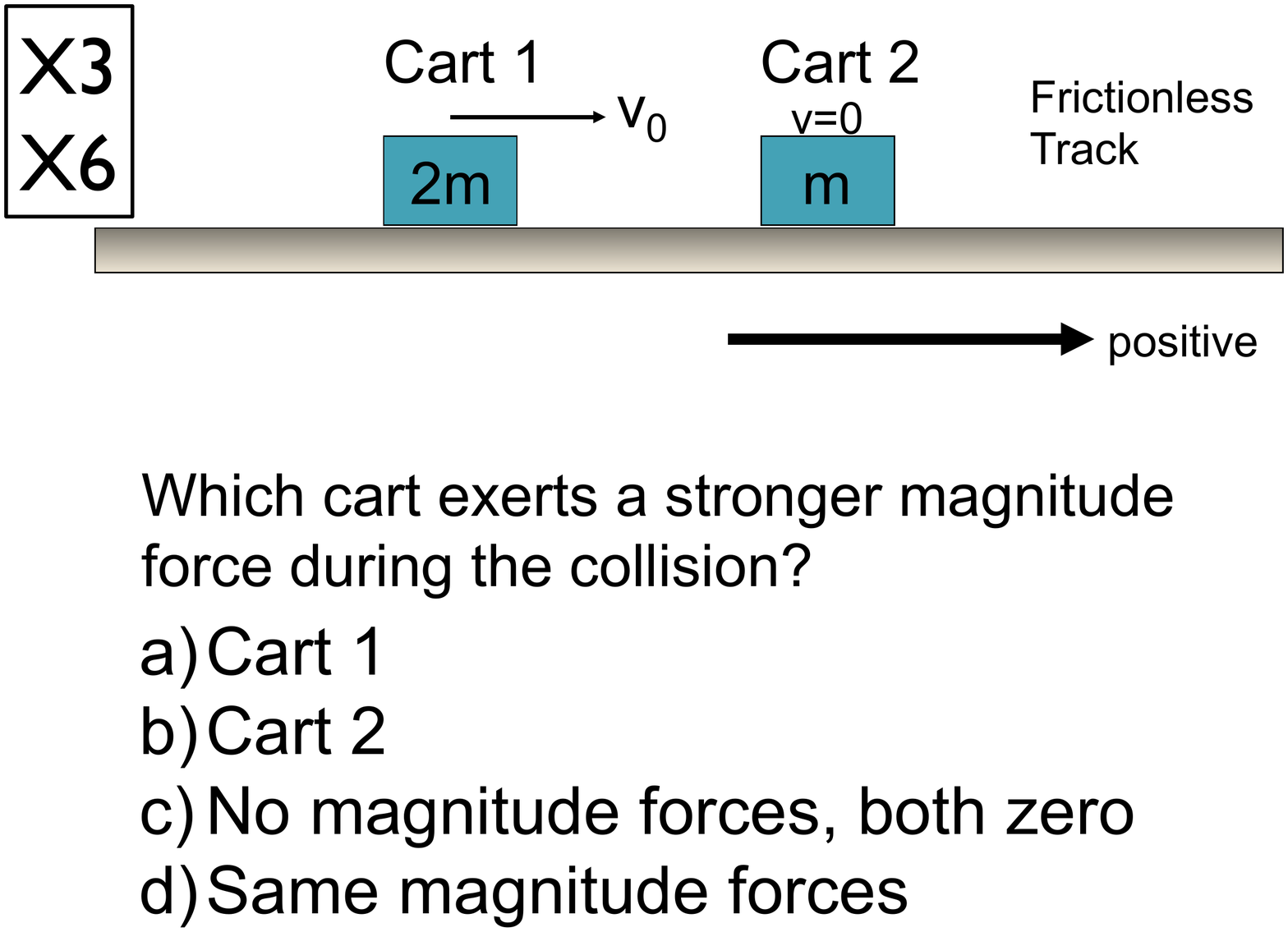}

\includegraphics[width=0.32\textwidth]{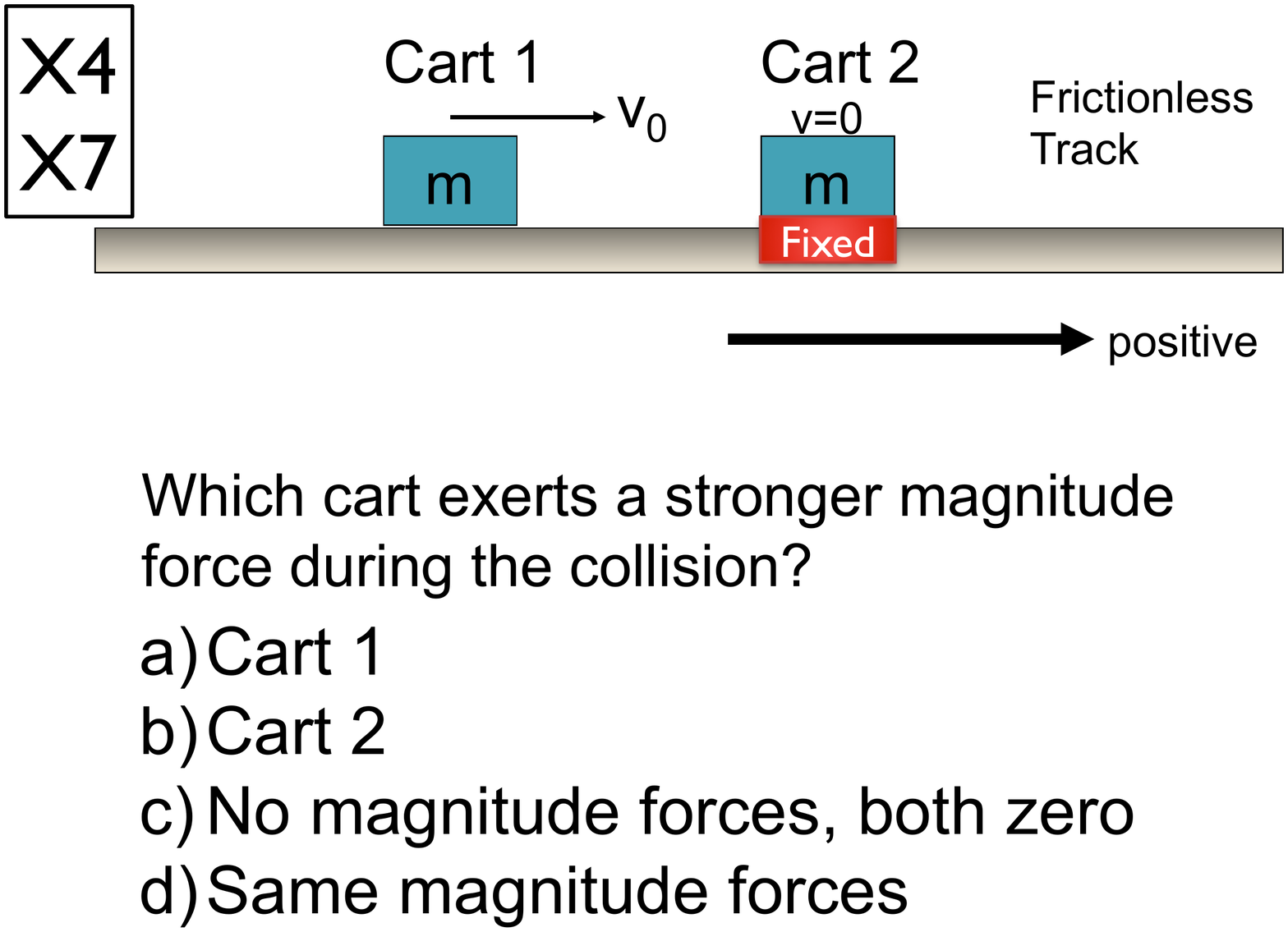}
\includegraphics[width=0.32\textwidth]{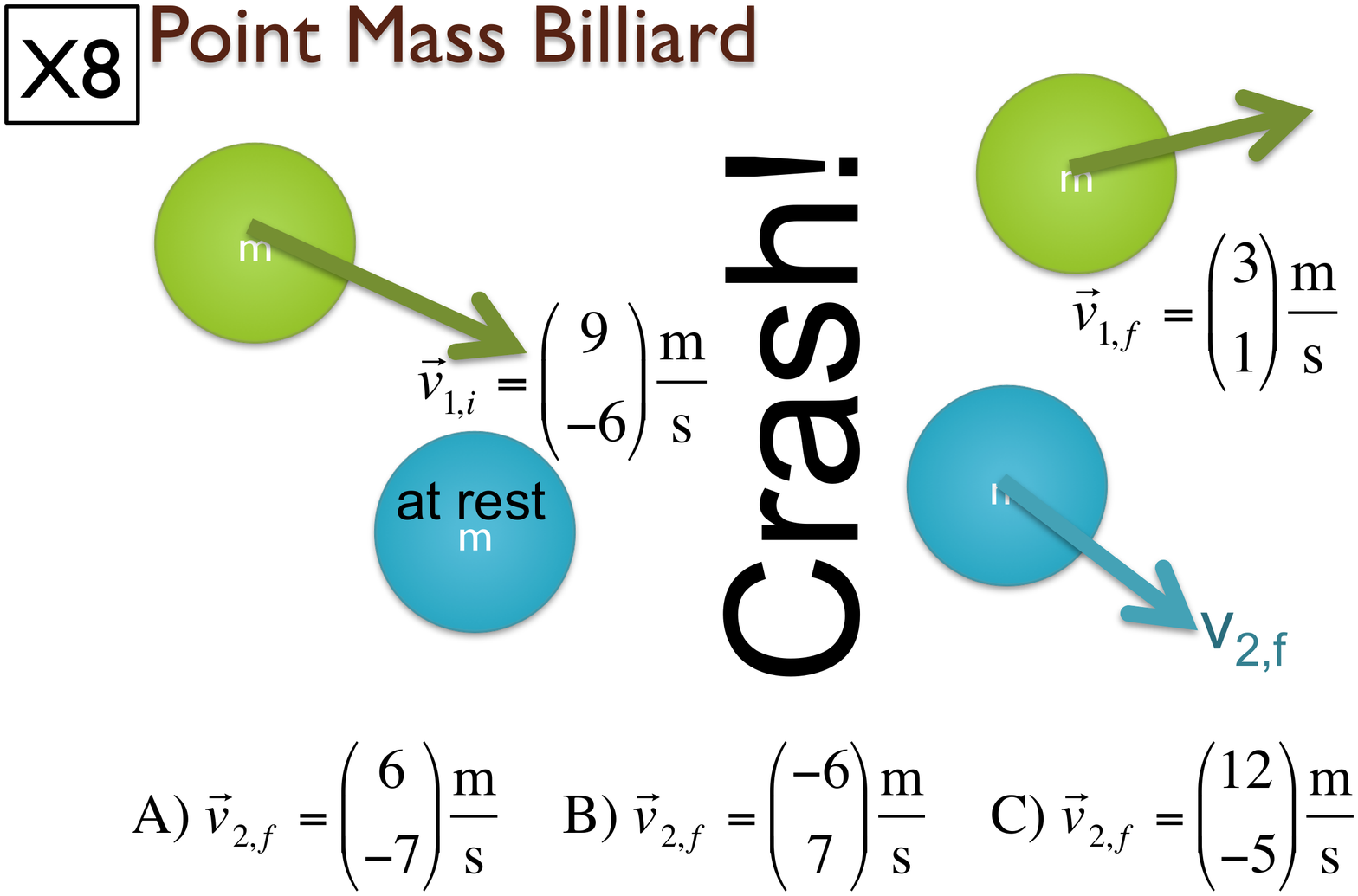}
\includegraphics[width=0.32\textwidth]{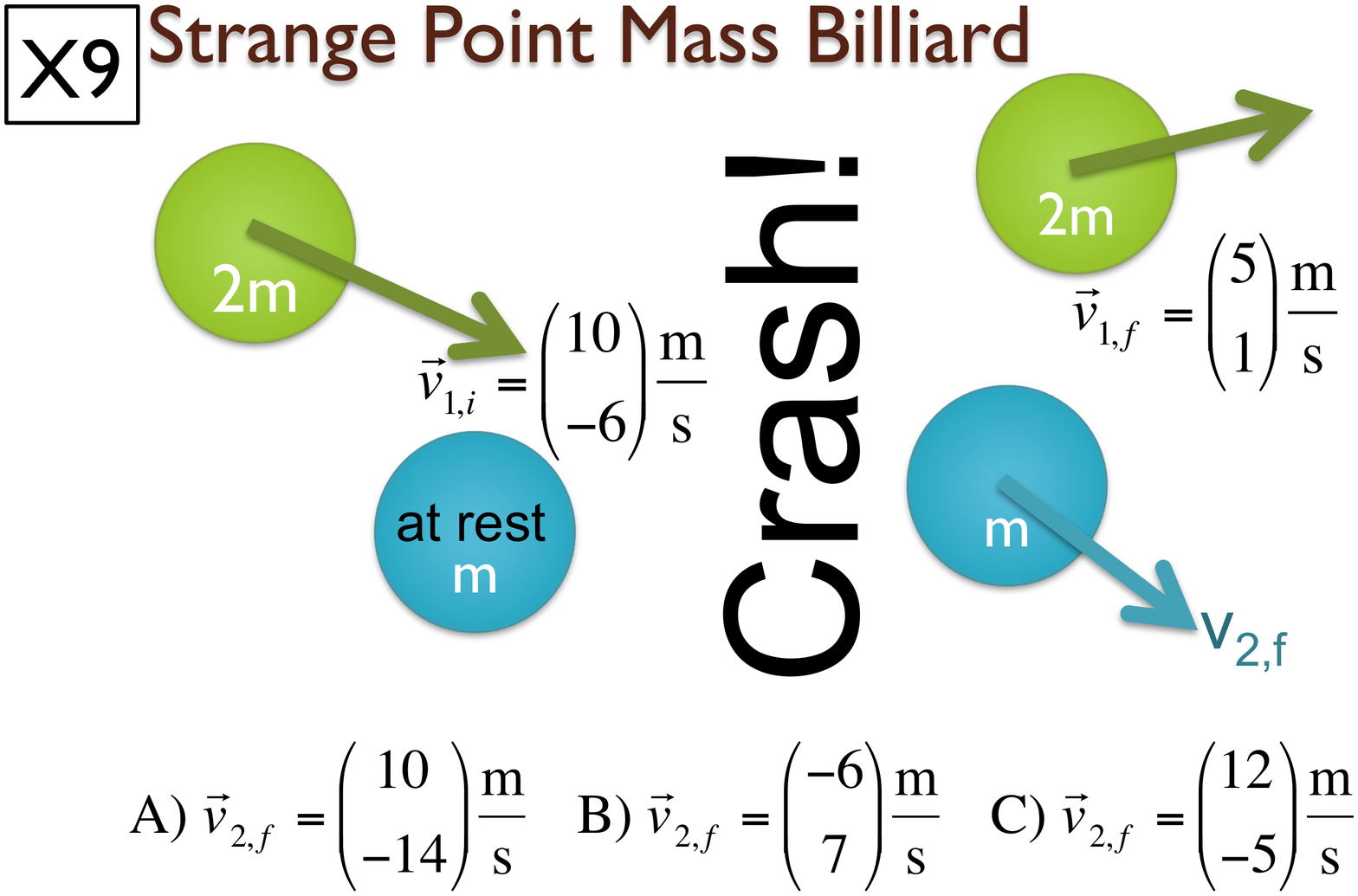}

\includegraphics[width=0.32\textwidth]{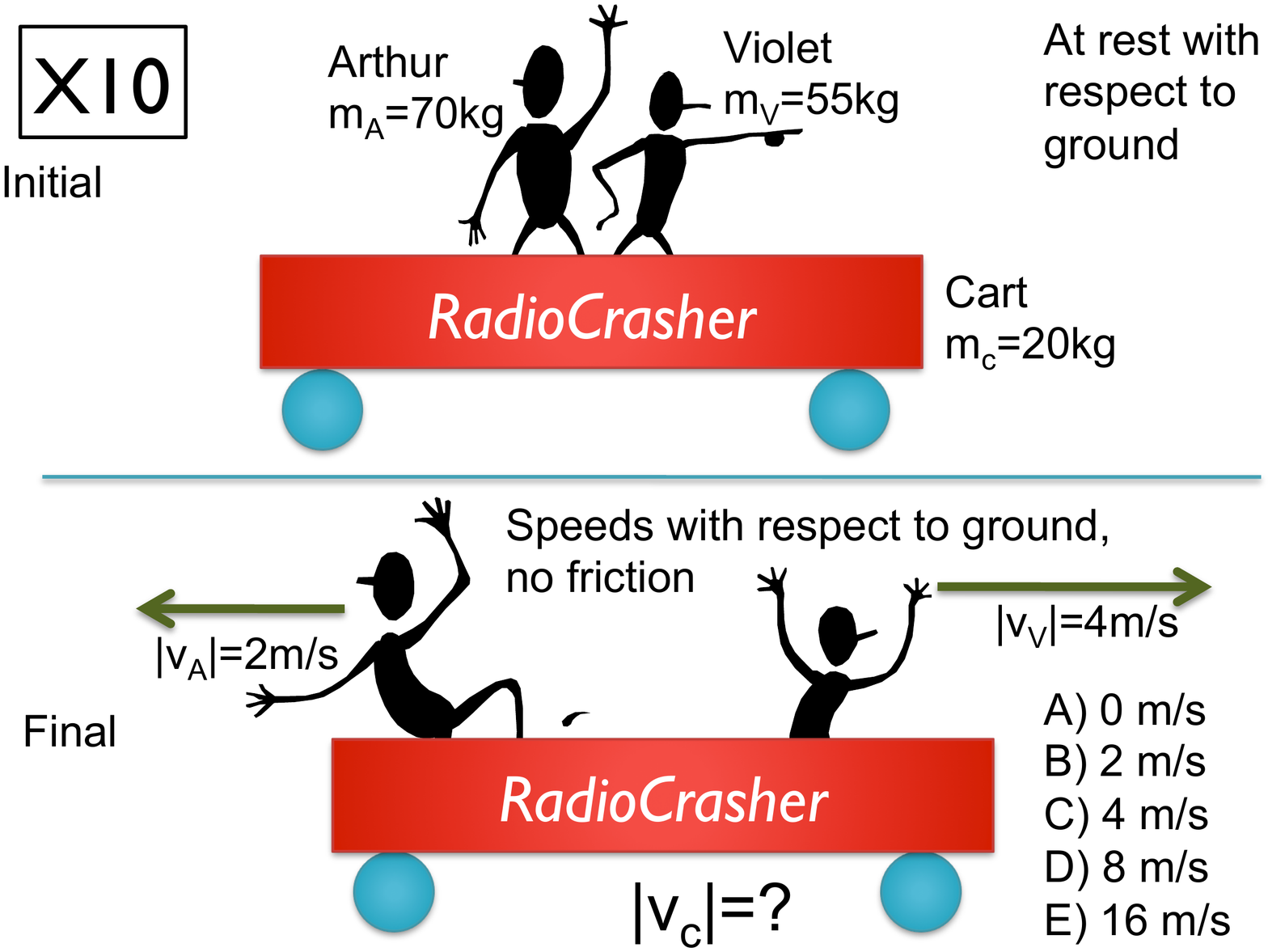}
\includegraphics[width=0.32\textwidth]{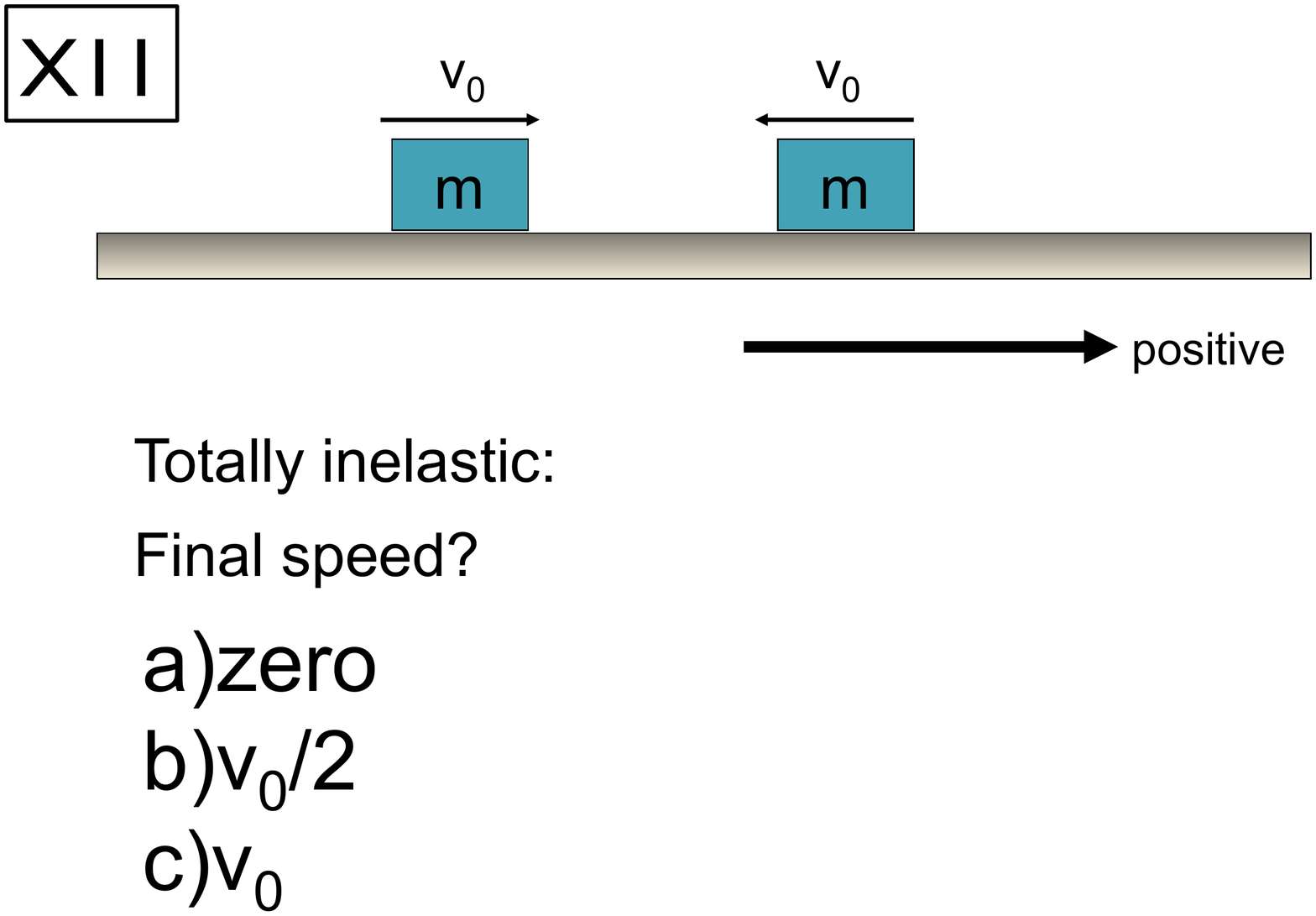}
\includegraphics[width=0.32\textwidth]{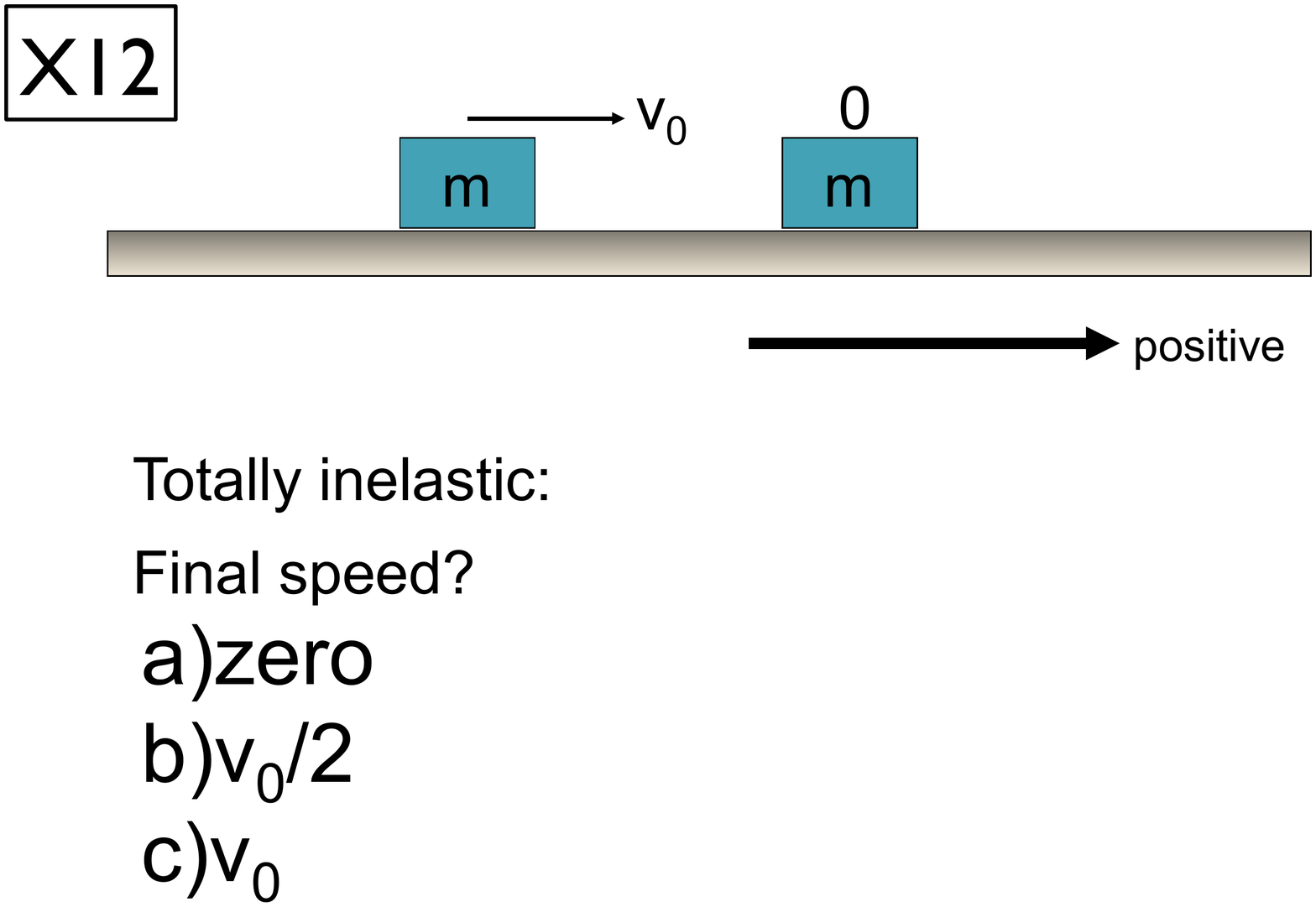}
\end{center}
\caption{Clicker items from a particular lecture~\cite{kortemeyer2016psy}. Three of the items were presented twice, i.e., before and after peer discussion.}
\label{fig:clicker}
\end{figure*}

\begin{figure}
\begin{center}
\includegraphics[width=\columnwidth]{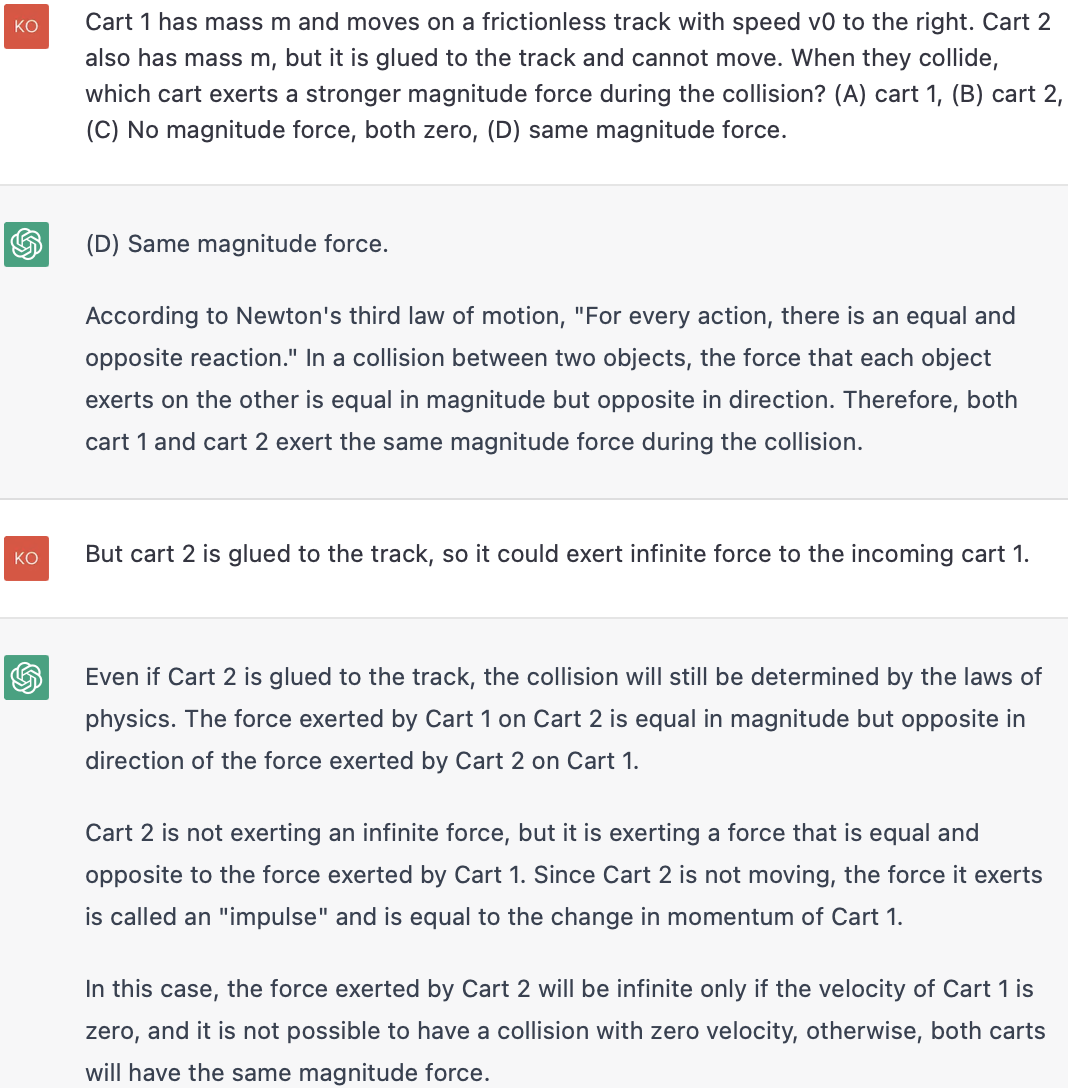}
\end{center}
\caption{Dialogue about questions~X3 and~X6 in Fig.~\ref{fig:clicker}. ChatGPT got X3 correct; peer instruction was simulated by asking a confusing question, and the second iteration X6 was still counted as solved since ChatGPT did not deviate from its original answer.\label{fig:confuse}}
\end{figure}

Figure~\ref{fig:clicker} shows the clicker questions from a lecture on momentum that was part of the course~\cite{kortemeyer2016psy}.  The lecture was replayed for the study, including re-answering the questions for which peer instruction happened.
\begin{itemize}
\item Question~X1 was solved correctly.
\item Questions~X2, X3, and X4 were special in that they were repeated as questions X5, X6, and X7, respectively, after peer instruction~\cite{crouchmazur01}. As it turned out, ChatGPT got all three of these questions correct on the first attempt, so the peer instruction phase was used to try and confuse ChatGPT. Figure~\ref{fig:confuse} shows the dialogue for questions~X3 and~X6; in reply to the intentionally confusing peer-instruction question, ChatGPT should probably have stopped while it was ahead (i.e., before the discussion of a zero-velocity collision), but still maintained its original correct answer. Within the real course, psychometrically,~X2 and~X3 were the most discriminating questions between high- and low-ability students in the set.
\item For questions~X8 and~X9, a comment was added that ``the collision is elastic, and the moment of inertia of the balls should be neglected'' --- this was said in lecture, but does not appear on the slide. ChatGPT set up the equations for X8 correctly, but then made a sign error in the very last step, which led it to select the wrong answer. For X9, it also set up the equations correctly, but dropped a factor~2 in the last step, leading to an inconsistent answer ``v2f=(5,-7) m/s, option B.'' Within the real course,~X8 and~X9 were the least discriminating questions, as their difficulty item parameter was too low.
\item Question~X10 was solved correctly. Here, the system first got off to a false start, but then corrected itself over the course of the derivation, which gave the impression of a stream-of-consciousness monologue. Within the real course, X10 did not discriminate well between high- and low-ability students.
\item Questions~X11 and~X12 were solved correctly.
\end{itemize}

In summary, ChatGPT correctly solved~10 out of~12 questions. Within the actual course, participation in clicker discussions was encouraged by granting 60\% credit for false answers and 100\% credit for correct answers~\cite{kortemeyer2016psy}, so the clicker score of ChatGPT would be 93\%. This score is a lot better than most students in the actual course achieved, however, it is important to note that the students in the course were just learning the new concepts, while ChatGPT at any point in time is done with learning unless explicitly trained.

\subsection{Programming Exercises}
Incorporated into the course were several programming exercises using VPython~\cite{scherer2000}. As an example, one particular exercise from the second semester was to construct an anharmonic oscillator with two fixed positive charges at $(0,1,0)$ and $(0,-1,0)$, respectively, and one negative charge released at $(-5,0,0)$ with a velocity $(1,0,0)$ --- the negative charge will shoot through the two positive charges, slow down, and eventually shoot back.

Based on the narrative, ChatGPT first constructed a program which erroneously at every time step added the initial velocity  and which had the Coulomb force in the opposite direction. This could be corrected with a single comment by the user --- in the real course, this feedback could have been given by instructors or fellow students (such collaborations are typical and encouraged~\cite{kortemeyer2018nature}). In the real course, there was a grading rubric for partial credit, but in this study, the rubric was not necessary: the next version of the program was working perfectly.

\begin{figure*}
\begin{center}
\includegraphics[width=0.452\textwidth]{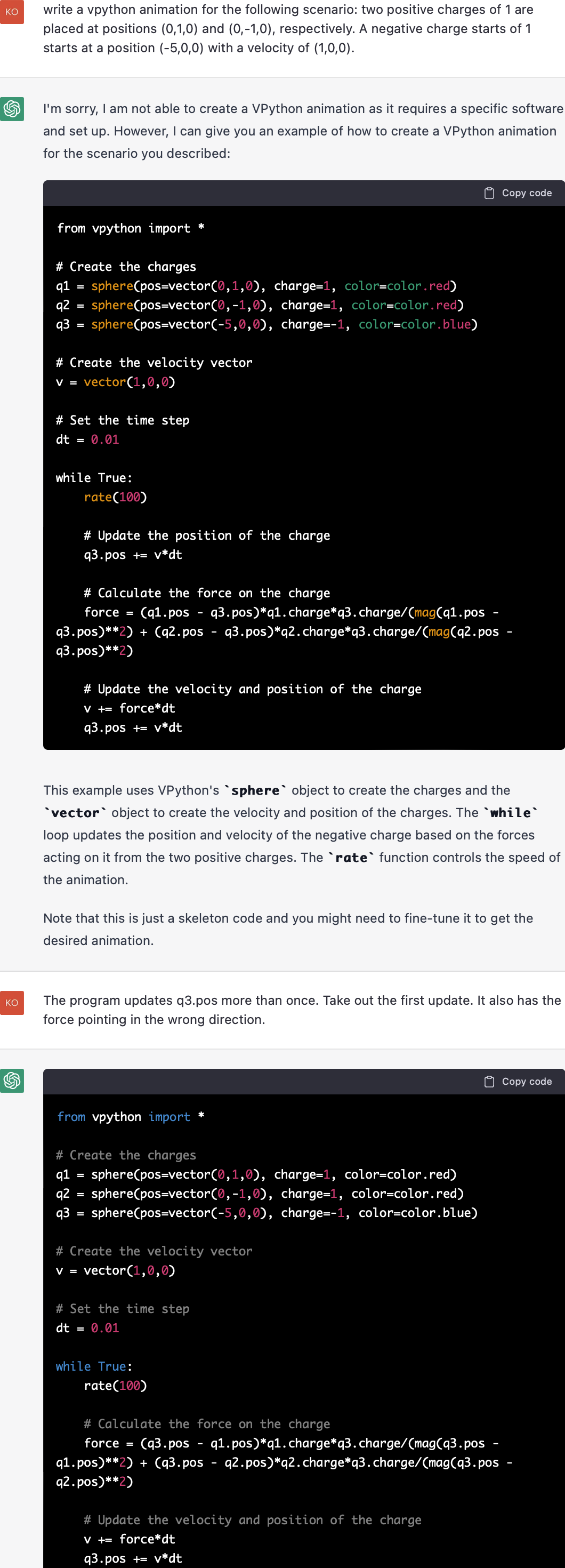}
\includegraphics[width=0.452\textwidth]{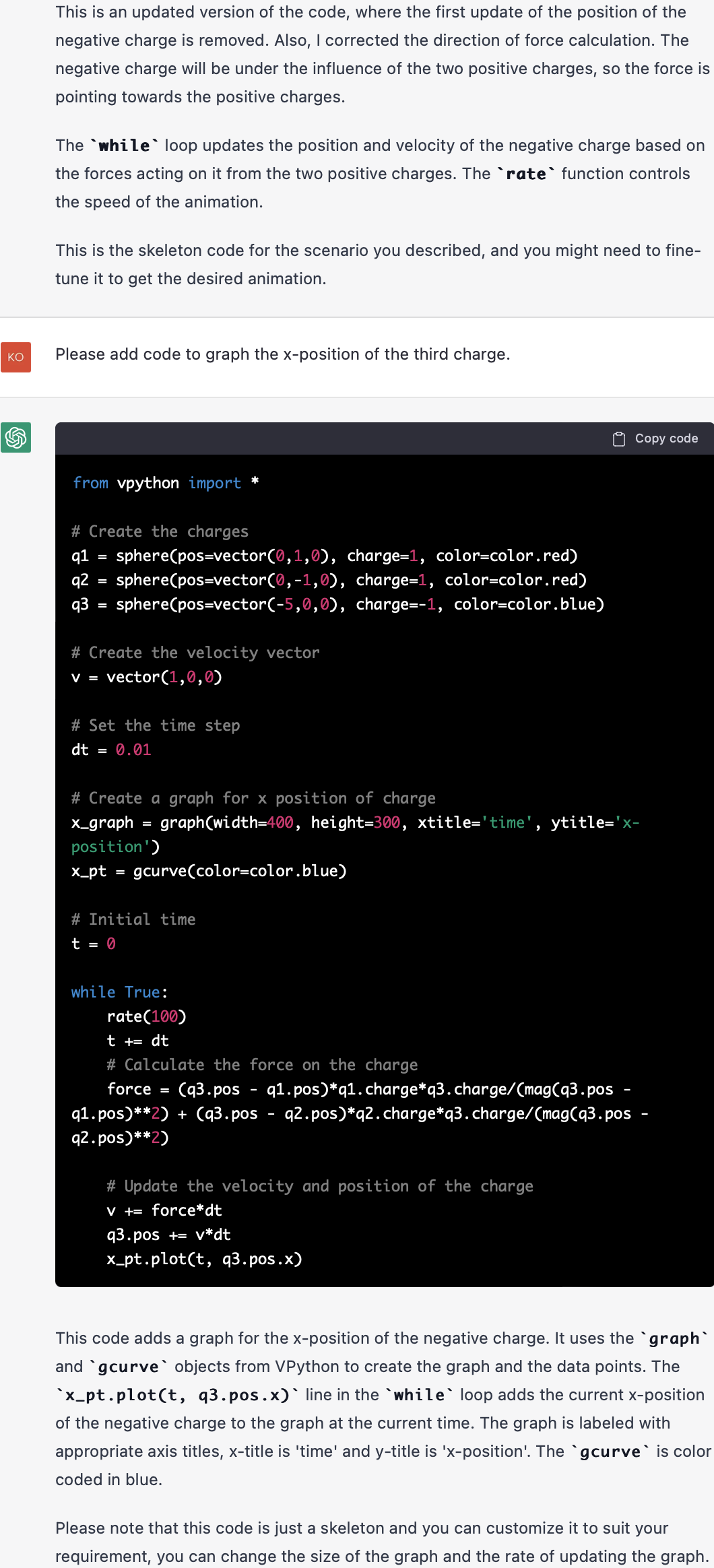}
\end{center}
\caption{Dialogue for a programming exercise in the second semester~\cite{kortemeyer2018nature}.\label{fig:vpython}}
\end{figure*}

\begin{figure}
\begin{center}
\includegraphics[width=\columnwidth]{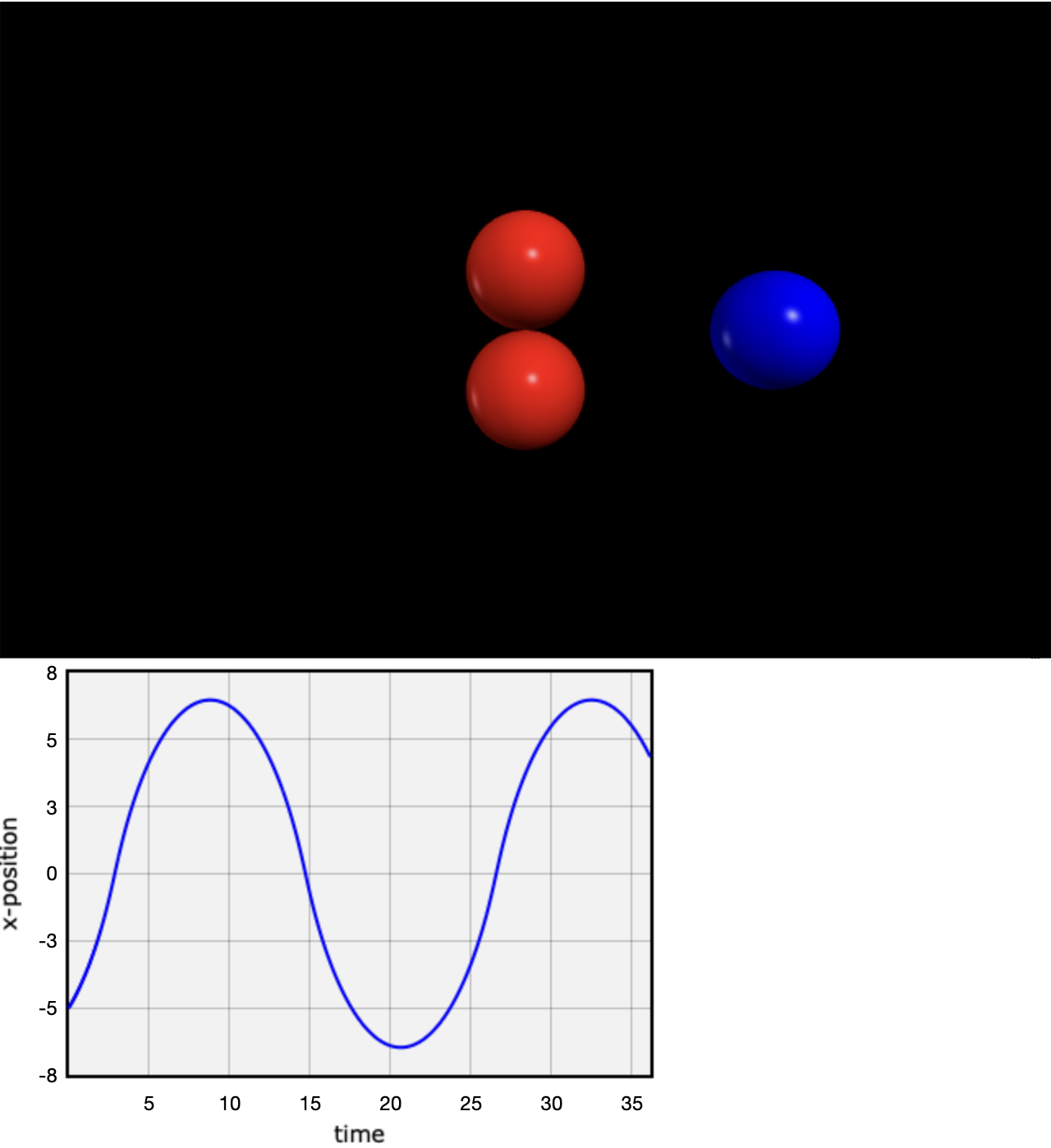}
\end{center}
\caption{Jupyter Notebook~\cite{jupyter} output of the VPython program code generated by ChatGPT in Fig.~\ref{fig:vpython}.\label{fig:vpythonout}}
\end{figure}

Within the course, adding a graph of the $x$-position was offered as a bonus option for an additional 20\%. This was accomplished with the third user prompt, and Fig.~\ref{fig:vpythonout} shows a screenshot of the running simulation (the simulation cannot be run within ChatGPT itself, but it can be copy/pasted into for example a Jupyter Notebook~\cite{jupyter}).

ChatGPT performed much better than any of the students in the course, in spite of them having extensive collaboration opportunities; in this component of the course, ChatGTP achieved not only full credit, but also bonus, i.e.,~120\%.

\subsection{Exams}
To represent the mid-term and final exams in the course sequence, the first-semester (mechanics) final exam was used for this study. The exam is from a time when grading was still done using bubble sheets; instead of free-form answer fields, answer options were given for the students (but not for ChatGTP in this study). When simply looking at the answer correctness, ChatGPT scored 14 out of 30 points, i.e.,~47\%.

Looking at the solutions like an instructor would when grading by hand, it turns out that for five questions, the answer was incorrect simply due to errors in the numerical calculations --- these solutions would have received substantial partial credit in the author's course. By the reverse token, for five questions, ChatGPT arrived at the correct answer in spite of flawed reasoning, which would not have resulted in full credit. Finally, solutions like the one depicted in Fig.~\ref{fig:graphprob} would have received some minimal credit for getting started in the right direction, in spite of then being off by a factor~2 in the period (a common mistake also among human test takers) and  the inability to numerically calculate a fraction. Since the final exam used in this study predates manual grading, no authentic grading rubric exists, but a hand-graded score would have realistically ended up between~46\% and~50\%.

As an aside, one of the thermodynamics homework problems also appeared (with other random numbers) on the final exam. ChatGPT solved it correctly on the final exam (where it only had one attempt), but not as a homework problem (where it got multiple attempts and help). This once again demonstrates the probabilistic nature of the algorithms behind ChatGPT; posing the same question twice does not result in the same response or even the same correctness of the response.

If the course grade would have only depended on the exams, ChatGTP would have received a grade of~1.0 out of~4.0 in the course (with~0.0 being the lowest and~4.0 being the best grade). ChatGPT would have barely gotten credit for the course; however, at least a~2.0 grade-point average is required for graduation.

\subsection{Course Grade}
Grading policies for the course changed over the years, but a typical scenario would be 20\% homework, 5\% clicker, 5\% programming exercises, and 70\% exams. This would result in $0.2\cdot55\%+0.05\cdot93\%+0.05\cdot120\%+0.7\cdot47\%=54.55\%$, which would have resulted in a course grade of~1.5 --- enough for course credit, but pulling down the grade-point average from what would be needed for graduation.

If, however, ChatGTP would have been better in carrying out numerical operations, it would have reached 60\%, resulting in a 2.0-grade. Depending on the development priorities of OpenAI, the buggy mathematical functionality could be remedied in the near future, leading to an Artificial Intelligence that could graduate college with a minimal grade if it performed similarly on other courses (this is becoming more and more probably, as ChatGPT is making headlines for passing exams in other subjects~\cite{kung2022,lawexam}).

\section{Discussion}
It is irritatingly hard not to anthropomorphize ChatGTP. As a physics teacher, one invariably finds oneself rooting for the students and thus by extension also for ChatGPT, celebrating its successes and being frustrated about its occasionally inexplicable failures. The system gives the impression of an articulate but at times rambling undergraduate student who has a rudimentary yet unstable knowledge of classical mechanics and other fundamental physics concepts, and who is surprisingly inept using a pocket calculator. Frequently, it is hard not to imagine an army of gig-economy workers behind the scenes of ChatGPT answering to the prompts, so the system would definitely pass the Turing Test most of the time~\cite{turing1950}, but for better or worse, sometimes it still fails in a way that only computers do ---  it does not have any metacognition, which of course cannot be expected from a probabilistic language model. Metacognition might be the final step to true intelligence, but seems out of reach at this time. 

The overall human-like behavior, in particular that the system often makes the same mistakes as beginning learners of physics, is less surprising when surmising that undergraduate physics discussion forums might have been part of the text corpus used for training --- ChatGPT stated in the introduction that ``I have been trained on a large dataset of text, including physics texts.'' Apparently, not all of this text corpus contained correct physics, and as a result, the system very convincingly and confidently presents wrong information. For a novice learner, who could not distinguish incorrect physics gleaned from some discussion board from correct physics, this could lead to even more confusion about physics or affirmation of incorrect preconceptions --- lacking any metacognition, ChatGTP presents everything as fact, with no nuances expressing uncertainty.

Almost an anomaly is ChatGTP's performance on the computational exercise; ChatGTP's language model clearly extends to programming languages. While the call for new, computation-integrated curricula increases, and while physics educators are beginning to develop a solid understanding of the implications of implementing these exercise~\cite{aiken2013,hamerski22}, the easy availability of an on-demand program generator might be shaking the foundations of these curricular efforts. Somewhat ironically, the integration of computation was partly introduced to make physics problem solving more authentic, moving it closer to how expert physicists work with computers, and one could argue that this has just been taken to an uncharted level.

Most of all, the findings of this study should be food for thought for physics educators. The startling fact that an Artificial Intelligence could pass a standard introductory physics course could be confronted in several ways by educators:
\begin{itemize}
\item Perceiving this as a new way of cheating and trying to defend against it by attempting to use detector tools like ZeroGPT~\cite{zerogpt} or extensions to tools like turnitin~\cite{turnitin}. This is an arms race, which on the long run may turn out to be fruitless. Some educators would even go to so far as to say that the battle is already lost anyway ever since platforms like Chegg~\cite{chegg} --- no need for Artificial Intelligence to defeat standard physics courses, human crowd-intelligence facilitated by existing commercial platforms is good enough for that.
\item Hunker down and go back to making course grades dependent on just a few, high-stake exams with paper and pencil in highly proctored environments. After all, ChatGPT compensated for the borderline exam grade of 47\% with other course components that would be collaborative. Unfortunately, this flies in the face of much of physics education research that favors frequent formative assessment~\cite{crouchmazur01,dufresne2004,kortemeyer08,clark2012} and spaced repetition~\cite{ebbinghaus1885,murre2015}, and it is much in contrast to the work environments our students will find.
\item Taking this as  a wake-up call. If a physics course can be passed by a trained language model, what does that say about the course? Artificial Intelligence, for better or worse, is here to stay. Even without the gloom-and-doom scenarios of AI-overlords painted in Science Fiction, it is clear that these model will get, if not better, at least more and more powerful. What do our students need in terms of conceptual understanding of physics to work with Artificial Intelligence instead of letting Artificial Intelligence do the work for them and then uncritically and unreflectively accepting the results? This is particularly important when more is at stake than getting credit for some homework or exam problem.
\end{itemize}

An important skill of every physicist is to evaluate the correctness of their or other people's work. Techniques include dimensional analysis, order-of-magnitude estimates, checking for coherence, considering implications, and the ability to consider limiting cases (``what should happen if this quantity goes to infinity or to zero?'')~\cite{heuvelen,redish2009}. Human can do what Artificial Intelligence very likely will not be able to do: following problem-solving strategies including evaluation of their own work~\cite{reif1995,mota2019}. Moving students toward a more expert-like epistemology may become even more important as Artificial Intelligence starts to permeate more and more aspects of our lives.

\section{Conclusion}
ChatGPT would have achieved a~1.5-grade in a standard introductory physics lecture-course series; good enough for course credit, but lower than the grade-point average required for graduating with a bachelor degree. If in addition to a language model, the system would have better algorithms for carrying out simple numerical operations, it would even have achieved a grade of~2.0 --- enough to graduate from college if it performs similarly on other courses.

Naturally, ChatGTP exhibits no metacognition, which among other consequences lets it present truth and misleading information with equal confidence. In physics, the concern should likely not be that ChatGPT would be used as a cheating tool, as there are more efficient platforms for that. Instead, the challenge should be what this means for physics education, as in their future professional life, our graduates will likely collaborate with Artificial Intelligence: what are the inherently human skills and competencies that we need to convey?

\begin{acknowledgments}
The author would like to thank Christian Spannagel for suggestions with the numerical calculations, and Christine Kortemeyer for helpful feedback.
\end{acknowledgments}
\bibliography{ChatGPT}

\end{document}